# Ultra-Narrow TaS₂ Nanoribbons

Jeffrey D. Cain[1,2,3†], Sehoon Oh[1,2†], Amin Azizi[1,3†], Scott Stonemeyer[1,2,3,4], Mehmet Dogan[1,2],

Markus Thiel[1], Peter Ercius[5], Marvin L. Cohen[1,2], and Alex Zettl[1,2,3*]

[1]*Department of Physics, University of California at Berkeley, Berkeley, CA 94720, USA*

[2]*Materials Sciences Division, Lawrence Berkeley National Laboratory, Berkeley, CA 94720, USA*

[3]*Kavli Energy NanoSciences Institute at the University of California at Berkeley and the Lawrence Berkeley National Laboratory, Berkeley, CA 94720, USA*

[4]*Department of Chemistry, University of California at Berkeley, Berkeley, CA 94720, USA*

[5]*The Molecular Foundry, Lawrence Berkeley National Laboratory, Berkeley, CA 94720 USA*

†These authors contributed equally

*e-mail: azettl@berkeley.edu

Imposing additional confinement in two-dimensional (2D) materials can yield further control over the associated electronic, optical, and topological properties. However, synthesis of ultra-narrow nanoribbons (NRs) remains a challenge, particularly for the transition metal dichalcogenides (TMDs), and synthesizing TMD NRs narrower than 50 nm has remained elusive. Here, we report the vapor-phase synthesis of ultra-narrow $TaS_2$ NRs. The NRs are grown within the hollow cavity of carbon nanotubes, thereby limiting their lateral dimensions and layer number, while simultaneously stabilizing them against the environment. The NRs reach the monolayer (ML) limit and exhibit widths as low as 2.5 nm. Atomic-resolution scanning transmission electron microscopy (STEM) reveals the detailed atomic structure of the ultra-narrow NRs and we observe a hitherto unseen atomic structure supermodulation phenomenon of ordered defect arrays within the NRs. First-principles calculations based on density functional theory (DFT) show the presence of flat bands, as well as edge- and boundary-localized states, and help identify the atomic configuration of the supermodulation. Nanotube-templated synthesis represents a unique, transferable, and broadly deployable route toward ultra-narrow TMD NR growth.

In step with the resurgent interest in 2D materials, there have been extensive efforts toward engineering additional levels of confinement, and thus lower dimensionality, in few- and monolayer van der Waals bonded 2D structures. The greatest successes have been achieved in the fabrication and synthesis of graphene nanoribbons (GNRs), where rational bottom-up synthesis has been accomplished using the self-assembly of molecular precursors (*1*, *2*). This has enabled the growth of GNRs with specific edge structures (*3*) and atomically-precise widths (*4*), as well as single GNR heterojunctions with engineered band alignment (*5*) and topology (*6*). Exciting physics is similarly predicted to arise when 2D transition metal dichalcogenides (TMDs) are further constrained towards one-dimension (1D), including the emergence of metal-insulator transitions (*7*), enhanced thermoelectric performance (*8*), ferromagnetism (*9*, *10*), and tunable band gaps (*7*).

However, the synthesis or fabrication of TMD NRs has lagged far behind that of GNRs both in terms of quality and width control. Past studies have relied upon top-down fabrication methods that require lithography and etching processes (*11*, *12*), which results in NRs with widths greater than 50 nm - too large to observe quantum confinement effects - and with a high degree of structural disorder (*13*). Molecular beam epitaxy has been used for the fabrication of ultra-narrow $MoSe_2$ NRs (*14*), but this method has little flexibility in implementation (*e.g.* substrate choice). Other bottom-up techniques, including chemical vapor deposition and vapor-liquid-solid growth, have recently been used to grow $MoS_2$ NRs (*15*, *16*), resulting in ribbons with average widths > 50 nm, again too large to access predicted quantum phenomena.

The templated growth of 1D nanomaterials using multi-walled carbon nanotubes (MWCNTs) has been successfully demonstrated for a variety of materials including elemental metals (*17*), halides (*18*), trichalcogenides (*19*, *20*), and molecular chains (*21*), where the synthesis generally results in nanowire- or chain-like structures. Here, we extend the method to ribbon-like morphologies, and demonstrate the growth of 2H-TaS$_2$ NRs. TaS$_2$ is a metallic TMD which hosts multiple charge-density-wave (CDW) phases (*22, 23*), a Mott insulator state (*24*) and possible quantum spin liquid (*25*), making it a unique material for the study of strongly correlated physics under extreme dimensional constraint. We achieve TaS$_2$ NRs with thicknesses down to the monolayer limit (typical layer numbers are between 1 and 3), widths as low as 2.5 nm, and lengths greater than 100 nm. The MWCNT sheath fully encapsulates the TaS$_2$ NR, protecting it from interaction with the environment. The growth produces NRs with controllable widths, clean NR surfaces and edges, and enables easy handling (*e.g.* solution-based processing) and subsequent characterization, all through a broadly deployable vapor-phase synthesis method. We investigate the structure and detailed atomic registry of NRs with transmission electron microscopy (TEM) and aberration-corrected annular dark-field scanning transmission electron microscopy (ADF-STEM). Furthermore, we observe and investigate previously unknown and unique periodic atomic superstructures defined by ordered defect arrays. First-principles calculations are used to elucidate the electronic structure of the NRs and atomic superstructure, revealing the presence of flat bands localized at the defect boundaries and edges.

TaS$_2$ NRs in the 2H phase are grown within MWCNT using chemical vapor transport (CVT), described in detail in the Methods section. In brief, MWCNTs are first opened at the end *via* oxidation at high temperature(*26*); the opened nanotubes are then coated onto the inner surface of a quartz ampule, which is filled with elemental Ta, S, and an iodine transport agent and sealed under vacuum. Synthesis is carried out in a gradient furnace in a manner similar to that used for single crystal TMD growth *via* CVT.

Fig. 1A shows a schematic of a monolayer $TaS_2$ NR within a carbon nanotube (for simplicity, a single-wall nanotube is shown), in both plan and end views. Figs. 1B-D show microscopy images of the as-synthesized material. Fig. 1B shows a plan view TEM image of a 5nm wide NR, while Fig. 1C shows an edge view TEM image of a 3-layer NR. The dashed yellow lines in Figs. 1B and 1C delineate the inner walls of the MWCNT. The strong contrast (dark lines are the $TaS_2$ atomic planes) in Fig. 1C is due to the edge-on orientation, resulting in greater sample thickness; the measured interlayer distance (~0.7 nm) matches well with that of bulk $TaS_2$ (0.7 nm) (*27*). Fig. 1D is an ADF-STEM image (plan view) of a monolayer $TaS_2$ NR approximately 4nm wide. The high crystallinity and uniform width of the NR is immediately apparent. It should be noted that Figs. 1B and C (TEM) have the reverse contrast of Fig. 1D (STEM).

We further explore the detailed atomic structure and phase of the $TaS_2$ NR using atomic-resolution ADF-STEM imaging. A pristine NR with width 3.8 nm is presented in Fig. 2A, where the hexagonal lattice characteristic of (monolayer) 2H-$TaS_2$ is clearly visible. A zoomed in portion of the ADF-STEM image of the NR in Fig. 1D is also shown in Fig. 2B. Fig. 2C is a simulated STEM images of a monolayer of 2H-$TaS_2$. Fig. 2D compares the line intensity profile along the dashed green box in the experimental image of Fig. 2B and that of a simulated STEM image of monolayer 2H-$TaS_2$ (Fig. 2C). This match suggests the experimental image is that of a monolayer 2H-$TaS_2$ NR, and further, is inconsistent with simulations performed on bi-layer $TaS_2$. The smallest NR width we have observed is 2.5nm, and lengths of order 100nm is not uncommon. The widest NR observed is 6 nm wide, and the average NR width is 3.8 nm. A histogram of NR widths is shown in Fig. S1. In addition to strictly planar morphologies, some NRs also show curling at the edges (Fig. 2E, plan view), presumably facilitated by the MWCNT confinement. The composition of the NRs is further supported by energy dispersive x-ray spectroscopy (EDS) (Figs. 2G-I), which shows, over the width and length of the NR, a reasonably uniform distribution of Ta and S atoms. The (wider, as expected) carbon contribution is from the MWCNT. The EDS spectrum collected from the $TaS_2$ NR also clearly shows peaks of Ta and S (Fig. S2); we also observe O (contamination), Si, and N (substrate), not shown in the displayed energy range. This further confirms the identity of the NR material.

We find that $TaS_2$ NRs can form not only with perfect atomic structure, but also with striking periodic atomic superstructure, an example of which is presented in Fig. 3. Fig. 3A shows an atomic-resolution ADF-STEM image of a 3.3 nm wide $TaS_2$ NR containing an ordered, "zig-zag" like superstructure characterized by triangular domains of ideal $TaS_2$ lattice interrupted by boundaries of low STEM contrast. The zig-zag structure is dramatically highlighted in a bandpass filtered version of the same image, presented in Fig. 3B. The period of the perturbation is ~9 unit cells.

As we discuss in more detail below, the 2H-TaS2 NRs considered here appear to support charge density waves (CDWs), but the CDW amplitudes are small and CDWs are *not* the origin of the dramatic zig-zag superstructure.  Rather, the superstructure arises from line defect arrays. The atomic structure of the defect arrays is calculated *via* first principles calculations (see below), and the relevant candidate structure is shown in Fig. 3C. The structure is characterized by zig-zag tracks of linearly-formed S vacancies. A STEM simulation of this structure is shown in Fig. 3D (STEM simulation details can be found in the Methods section) for comparison with the experimental image in Fig. 3E. Intensity line profiles across the boundary (along the dashed green box) in both the simulated (Fig. 3F) and experimental images (Fig. 3G) match, showing a Ta-Ta distance of ~0.47 nm at the boundary versus ~0.32 nm within the ideal $TaS_2$ lattice.

We expand on our first-principles calculations based on DFT. We investigate the atomic and electronic structures of $TaS_2$ bulk and ML, and the obtained atomic and electronic structures of the 2H-bulk and ML configurations (Fig. S3) are consistent with other studies (*28, 29*). We investigate the atomic and electronic structure of NRs with various width considering possible CDW distortions and other structural defects. First, we start with the NRs without structural defects (*i.e.* no vacancies, substitutions, or adatoms). We construct candidate structures with various periodicity for considering possible CDW type distortions. The atomic positions of all the constructed structures are relaxed by minimizing the total energy. The obtained atomic and electronic structures of a NR ($W = 2.99$ nm) are shown in Figs. 4A-C. Because of the metallicity and 1D nature, CDW-type distortions and corresponding partial gap openings are found in all the structures with various width. Fig. 4A incorporates the CDW distortions, but the amplitude of the distortions is too small to be readily apparent, and furthermore, the CDW distortions are not commensurate with the experimentally observed zig-zag pattern. Therefore, we exclude the possibility that the zig-zag superstructure originates from CDW-type distortions without structural defects.

We therefore construct and optimize numerous new candidate structures of the NRs with various types of structural defects for the zig-zag boundaries. The atomic positions of all the constructed candidate structures are relaxed again by minimizing the total energy. Among the various candidate structures, we find a defect structure of NRs with various widths that is energetically favored compared to other candidate structures and matches well with the experimental STEM data. After careful analysis of the energetics of all the obtained structures (described below) and comparison with the STEM data, we conclude that the zigzag boundaries are linearly-formed S vacancies in the NRs. As described above, Fig. 3 shows the obtained atomic structures of a 2H-TaS$_2$ NR with $W$=3.34 nm, the experimental STEM data, and the STEM simulation using the structures obtained by DFT for comparison, which agree well with each other. Figs. 4D-H show the atomic and electronic structures of a NR with the zigzag defect array ($W$ = 3.08 nm) in vacuum. In Fig. 4D, the atomic structure is presented with the zigzag defect array denoted by $L_1$ and $L_2$, and the mirror planes denoted by $M_1$ and $M_2$. Figs. 4E-F show the band structure unfolded with respect to the unit-cell of the primitive 2H-ML and the projected density of states (PDOS), respectively. A partial gap opening around the K-point is observed as shown in Fig. 4E and there are two flat bands near the Fermi energy, $E_F$, (0.020 eV and 0.007 eV below $E_F$), denoted as $\psi_1$ and $\psi_2$, respectively, in Figs. 4E-F. The real-space wave function of the flat bands reveals that they are localized edge states as shown in Figs. 4G-H. The state $\psi_1$ is localized at the lower edge and has odd parity with respect to the mirror symmetries $M_1$ and $M_2$, while $\psi_2$ is at the upper edge and has even parity with respect to the mirror symmetries as shown in Figs. 4G-H. A similar analysis is also performed on pristine and defective NRs of width 3.29 nm, the results of which are shown in Fig. S4.

To analyze the stability of the NRs with various numbers of atoms in the unit-cell, we calculate the Gibbs free energy of formation, $\delta G$, of the obtained structures, which is defined as $\delta G = E_{NR} + n_{Ta}\mu_{Ta} + n_S\mu_S$, where $E_{NR}$ is the total energy per atom of the NR, $n_{Ta}$ and $n_S$ are the mole fractions of Ta and S atoms, respectively, and $\mu_{Ta}$ and $\mu_S$ are the chemical potentials of Ta and S, respectively. We choose $\mu_{Ta}$ and $\mu_S$ as the binding energies per atom of the α-Ta bulk and crown-shaped S$_8$ molecule, respectively. We find an interesting tendency that the structures with more S vacancies become more stable. For a quantitative analysis, we measure the ratio $R$ of the number of Ta atoms to that of S atoms, which is defined as $R = n_S / n_{Ta}$. The NRs without structural defects have $R = 2.14 \sim 2.20 > R_{bulk} = 2.0$, because the edges are S-terminated, and the confined nature of the NR. The presence of S vacancies reduces $R$, and the stable calculated structures agreeing well with the experimental data have $R = 2.00\text{~}2.01$ (Tab. 1). We note that other kinds of defect ordering, such as Ta adatoms, can also reduce $R$ and thereby enhance structural stability. We speculate that stabilization of the structure afforded by driving $R \rightarrow 2$ is the main driving force of the formation of the zigzag defects with S vacancies.

It is worth emphasizing the following: First, the calculated size of the partial gaps due to CDW-type distortions is ~0.1 eV, implying that the CDW distortions would persist at room temperature. Second, the encapsulation of the NRs in CNTs does not alter the electronic structures of the NRs except for the slight changes of $E_F$ due to charge transfer between NRs and CNTs. The calculated charge transfer $q$ from CNTs to 2H- NRs is ~0.02 $e$ per Ta atom, where $e$ is the electron charge. Third, no significant magnetic moment is found in all of the fully magnetic calculations with spin–orbit interaction. Last, the presence of the time-reversal symmetry and mirror symmetry in 2H- phase indicates the potential for interesting topological properties in these NRs.

In summary, we have demonstrated a method that enables the synthesis of ultra-narrow TMD NRs *via* a simple and broadly deployable gas phase process, using MWCNTs to template their growth. Our method results in 2H-TaS$_2$ NRs with widths below 3 nm, and lengths greater than 100 nm, while reaching the ML limit. Further, the nanoconfined growth of the NRs results in the formation of ordered arrays of linear defects in the TaS$_2$ lattice. DFT calculations reveal the characteristics of the defect arrays with spatial symmetries and electronic structures with localized edge and boundary states. The reported nanotube templated growth represents a versatile platform for the synthesis of ultra-narrow TMD NRs and the exploration of the materials under constraint in multiple dimensions.

**Acknowledgements**

This work was primarily funded by the U.S. Department of Energy, Office of Science, Office of Basic Energy Sciences, Materials Sciences and Engineering Division, under Contract No. DE-AC02-05- CH11231 within the sp2-Bonded Materials Program (KC2207) and the Nanomachines Program which provided for synthesis of the chains, TEM structural characterization, and theoretical modeling of relaxed structure of the NRs. The elemental mapping work was funded by the U.S. Department of Energy, Office of Science, Office of Basic Energy Sciences, Materials Sciences and Engineering Division, under Contract No. DE-AC02-05- CH11231 within the van der Waals Heterostructures Program (KCWF16). Work at the Molecular Foundry (TEAM 0.5 characterization) was supported by the Office of Science, Office of Basic Energy Sciences, of the U.S. Department of Energy under Contract No. DE-AC02-05-CH11231. Support was also provided by the National Science Foundation under Grant No. DMR-1807233 which provided for preparation of opened nanotubes and Grant No. DMR 1926004 which provided for theoretical calculations of the electronic band structure of the materials. Computational resources were provided by the DOE at Lawrence Berkeley National Laboratory's NERSC facility and the NSF through XSEDE resources at NICS.

**Author Contributions**

J. D. C. and A. Z. conceived of the idea; J. D. C. and M. T. synthesized the materials; J. D. C., A. A., S. S. and P. E. performed electron microscopy data acquisition and analysis. S. O. and M.D. carried out density functional calculations. S. O. and M. D. did the theoretical analysis. A. Z. and M. L. C supervised the project; J. D. C, A.A., and S. O. wrote the manuscript with input from all authors.

**Competing interests**

The authors declare no competing financial interest.

**Additional Information**

Correspondence and requests for materials should be addressed to A. Z.

Email: azettl@berkeley.edu

**References**

1.  P. B. Bennett, Z. Pedramrazi, A. Madani, Y.-C. Chen, D. G. de Oteyza, C. Chen, F. R. Fischer, M. F. Crommie, J. Bokor, Bottom-up graphene nanoribbon field-effect transistors. *Appl. Phys. Lett.* **103**, 253114 (2013).

2.  Y.-C. Chen, D. G. de Oteyza, Z. Pedramrazi, C. Chen, F. R. Fischer, M. F. Crommie, Tuning the Band Gap of Graphene Nanoribbons Synthesized from Molecular Precursors. **22**, 19 (2013).

3.  C. Tao, L. Jiao, O. V. Yazyev, Y.-C. Chen, J. Feng, X. Zhang, R. B. Capaz, J. M. Tour, A. Zettl, S. G. Louie, H. Dai, M. F. Crommie, Spatially resolving edge states of chiral graphene nanoribbons. *Nat. Phys.* **7**, 616–620 (2011).

4.  J. Cai, C. A. Pignedoli, L. Talirz, P. Ruffieux, H. Söde, L. Liang, V. Meunier, R. Berger, R. Li, X. Feng, K. Müllen, R. Fasel, Graphene nanoribbon heterojunctions. *Nat. Nanotechnol.* **9**, 896–900 (2014).

5.  D. J. Rizzo, G. Veber, T. Cao, C. Bronner, T. Chen, F. Zhao, H. Rodriguez, S. G. Louie, M. F. Crommie, F. R. Fischer, Topological band engineering of graphene nanoribbons. *Nature*. **560**, 204–208 (2018).

6.  G. D. Nguyen, H. Z. Tsai, A. A. Omrani, T. Marangoni, M. Wu, D. J. Rizzo, G. F. Rodgers, R. R. Cloke, R. A. Durr, Y. Sakai, F. Liou, A. S. Aikawa, J. R. Chelikowsky, S. G. Louie, F. R. Fischer, M. F. Crommie, Atomically precise graphene nanoribbon heterojunctions from a single molecular precursor. *Nat. Nanotechnol.* **12**, 1077–1082 (2017).

7.  K. Dolui, C. Das Pemmaraju, S. Sanvito, Electric Field Effects on Armchair MoS 2 Nanoribbons (2012), doi:10.1021/nn301505x.

8.  Z. Zhang, Y. Xie, Q. Peng, Y. Chen, A theoretical prediction of super high-performance thermoelectric materials based on MoS2/WS2 hybrid nanoribbons. *Sci. Rep.* **6**, 21639 (2016).

9.  A. R. Botello-Méndez, F. López-Urías, M. Terrones, H. Terrones, Metallic and ferromagnetic edges in molybdenum disulfide nanoribbons. *Nanotechnology*. **20**, 325703 (2009).

10. D. Davelou, G. Kopidakis, E. Kaxiras, I. N. Remediakis, Nanoribbon edges of transition-metal dichalcogenides: Stability and electronic properties. *Phys. Rev. B*. **96**, 165436

(2017).

11.  Q. Li, J. T. Newberg, E. C. Walter, J. C. Hemminger, R. M. Penner, Polycrystalline Molybdenum Disulfide (2H−MoS 2 ) Nano-and Microribbons by Electrochemical/Chemical Synthesis. **19**, 47 (2019).

12.  Y. Pak, Y. Kim, N. Lim, J.-W. Min, W. Park, W. Kim, Y. Jeong, H. Kim, K. Kim, S. Mitra, B. Xin, T.-W. Kim, I. S. Roqan, B. Cho, G.-Y. Jung, Scalable integration of periodically aligned 2D-MoS 2 nanoribbon array. *APL Mater.* **6**, 076102 (2018).

13.  Y. Li, E. C. Moy, A. A. Murthy, S. Hao, J. D. Cain, E. D. Hanson, J. G. DiStefano, W. H. Chae, Q. Li, C. Wolverton, X. Chen, V. P. Dravid, Large-Scale Fabrication of MoS 2 Ribbons and Their Light-Induced Electronic/Thermal Properties: Dichotomies in the Structural and Defect Engineering. *Adv. Funct. Mater.* **28**, 1704863 (2018).

14.  F. Cheng, H. Xu, W. Xu, P. Zhou, J. Martin, K. P. Loh, Controlled Growth of 1D MoSe 2 Nanoribbons with Spatially Modulated Edge States. *Nano Lett.* **17**, 1116–1120 (2017).

15.  S. Li, Y.-C. Lin, W. Zhao, J. Wu, Z. Wang, Z. Hu, Y. Shen, D.-M. Tang, J. Wang, Q. Zhang, H. Zhu, L. Chu, W. Zhao, C. Liu, Z. Sun, T. Taniguchi, M. Osada, W. Chen, Q.-H. Xu, A. Thye, S. Wee, K. Suenaga, F. Ding, G. Eda, Vapour-liquid-solid growth of monolayer MoS 2 nanoribbons. *Nat. Mater.* **17**, 535–542 (2018).

16.  T. Chowdhury, J. Kim, E. C. Sadler, C. Li, S. W. Lee, K. Jo, W. Xu, D. H. Gracias, N. V. Drichko, D. Jariwala, T. H. Brintlinger, T. Mueller, H. G. Park, T. J. Kempa, Substrate-directed synthesis of MoS2 nanocrystals with tunable dimensionality and optical properties. *Nat. Nanotechnol.* **15**, 29–34 (2020).

17.  T. Pham, A. Fathalizadeh, B. Shevitski, S. Turner, S. Aloni, A. Zettl, A Universal Wet-Chemistry Route to Metal Filling of Boron Nitride Nanotubes. *Nano Lett.* **16**, 320–325 (2016).

18.  E. Philp, J. Sloan, A. I. Kirkland, R. R. Meyer, S. Friedrichs, J. L. Hutchison, M. L. H. Green, An encapsulated helical one-dimensional cobalt iodide nanostructure. *Nat. Mater.* **2**, 788–791 (2003).

19.  T. Pham, S. Oh, P. Stetz, S. Onishi, C. Kisielowski, M. L. Cohen, A. Zettl, Torsional instability in the single-chain limit of a transition metal trichalcogenide. *Science (80-. ).* **361**, 263–266 (2018).

20.  S. Meyer, T. Pham, S. Oh, P. Ercius, C. Kisielowski, M. L. Cohen, A. Zettl, "Metal-

insulator transition in quasi-one-dimensional HfTe3 in the few-chain limit," (available at https://arxiv.org/ftp/arxiv/papers/1903/1903.00464.pdf).

21.    T. Pham, S. Oh, S. Stonemeyer, B. Shevitski, J. D. Cain, C. Song, P. Ercius, M. L. Cohen, A. Zettl, Emergence of Topologically Nontrivial Spin-Polarized States in a Segmented Linear Chain. *Phys. Rev. Lett.* **124**, 206403 (2020).

22.    A. W. Tsen, R. Hovden, D. Wang, Y. D. Kim, J. Okamoto, K. A. Spoth, Y. Liu, W. Lu, Y. Sun, J. C. Hone, L. F. Kourkoutis, P. Kim, A. N. Pasupathy, Structure and control of charge density waves in two-dimensional 1T-TaS2. *Proc. Natl. Acad. Sci. U. S. A.* **112**, 15054–9 (2015).

23.    B. Burk, R. E. Thomson, J. Clarke, A. Zettl, Surface and Bulk Charge Density Wave Structure in 1 T-TaS2. *Science.* **257**, 362–4 (1992).

24.    B. Sipos, A. F. Kusmartseva, A. Akrap, H. Berger, L. Forr´o, F. Forr´o, E. Tutiš, T. Tutiš, From Mott state to superconductivity in 1T-TaS 2 (2008), doi:10.1038/nmat2318.

25.    K. T. Law, P. A. Lee, 1T-TaS2 as a quantum spin liquid. *Proc. Natl. Acad. Sci. U. S. A.* **114**, 6996–7000 (2017).

26.    P. M. Ajayan, T. W. Ebbesen, T. Ichihashi, S. Iijima, K. Tanigaki, H. Hiura, Opening carbon nanotubes with oxygen and implications for filling. *Nature.* **362**, 522–525 (1993).

27.    R. Hovden, A. W. Tsen, P. Liu, B. H. Savitzky, I. El Baggari, Y. Liu, W. Lu, Y. Sun, P. Kim, A. N. Pasupathy, L. F. Kourkoutis, Atomic lattice disorder in charge-density-wave phases of exfoliated dichalcogenides (1T-TaS2). *Proc. Natl. Acad. Sci. U. S. A.* **113**, 11420–11424 (2016).

28.    P. Lazar, J. Martincová, M. Otyepka, Structure, dynamical stability, and electronic properties of phases in TaS2 from a high-level quantum mechanical calculation. *Phys. Rev. B - Condens. Matter Mater. Phys.* **92**, 224104 (2015).

29.    C. E. Sanders, M. Dendzik, A. S. Ngankeu, A. Eich, A. Bruix, M. Bianchi, J. A. Miwa, B. Hammer, A. A. Khajetoorians, P. Hofmann, Crystalline and electronic structure of single-layer TaS2. *Phys. Rev. B.* **94**, 081404 (2016).

**Figures and Tables**

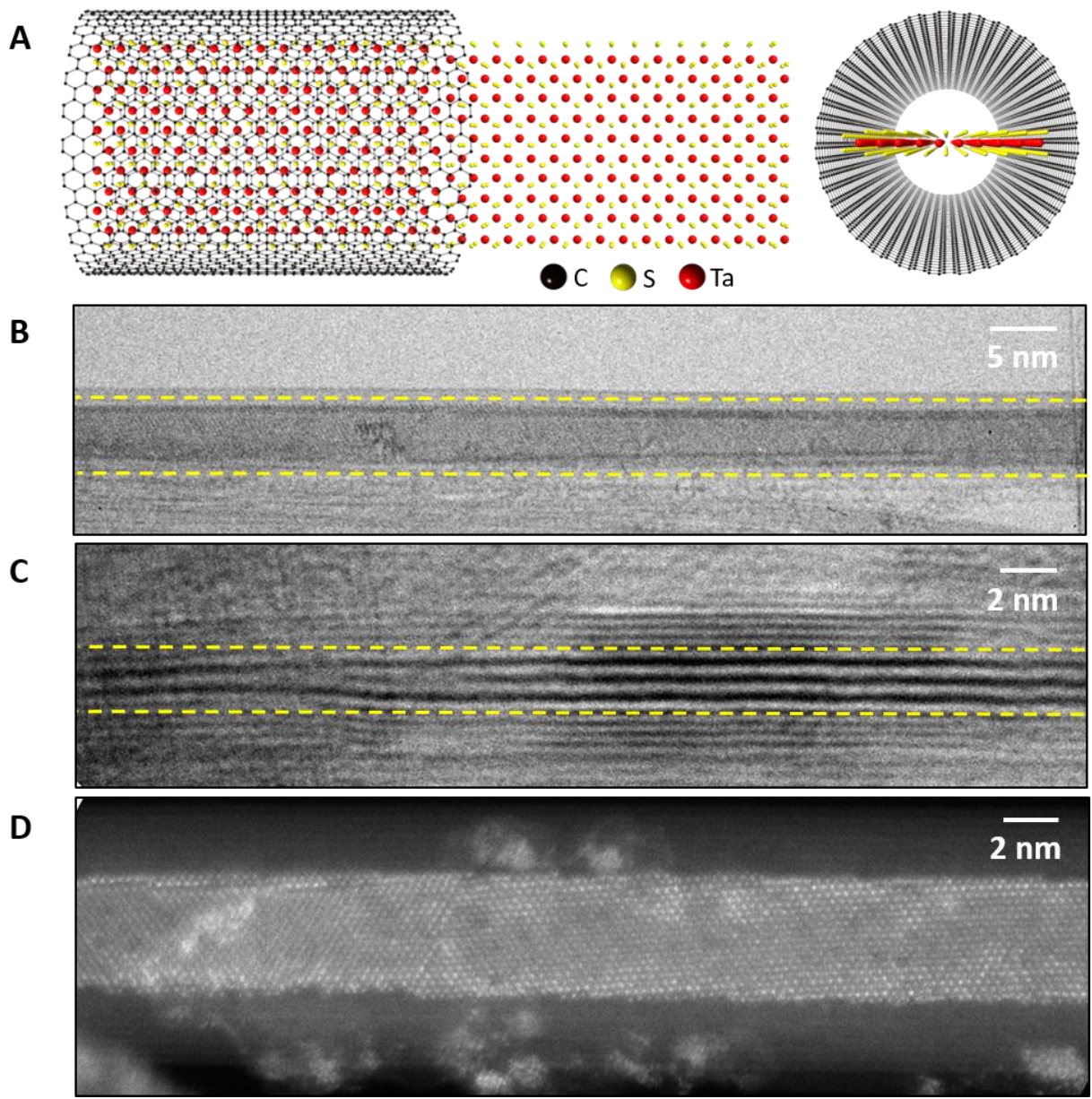

**Fig. 1 Imaging of ultra-narrow TMD nanoribbons.** (**A**) Schematic of $TaS_2$ nanoribbons templated by carbon nanotubes. (**B**) Plan view TEM image of 5 nm wide $TaS_2$ nanoribbon. (**C**) Side-view TEM image of a multilayer 2H-$TaS_2$ nanoribbon. (**D**) High-resolution ADF-STEM image of a pristine $TaS_2$ nanoribbon with ~4nm width.

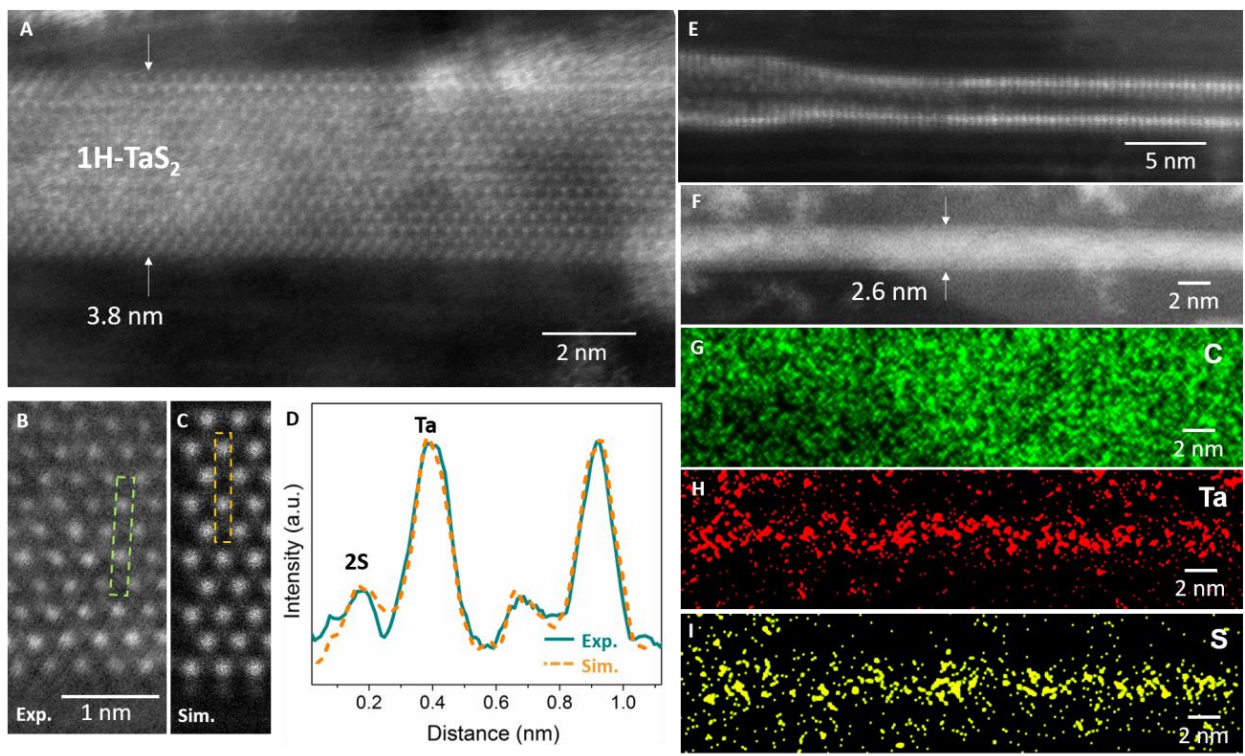

**Fig. 2. Atomic-resolution imaging of ultra-narrow, monolayer TaS₂ nanoribbons.** (**A**) and (**B**) Atomic resolution ADF-STEM images of 2H-TaS₂ nanoribbons. (**C**) STEM simulation of 2H-TaS₂ monolayer nanoribbon (**D**) Line scan intensity profile comparison of the experimental (Green) and simulated (orange) STEM images. (**E**) Side-view ADF-STEM image showing nanoribbon folding. (**F-I**) ADF-STEM reference image and EDS maps of C, Ta, S, respectively.

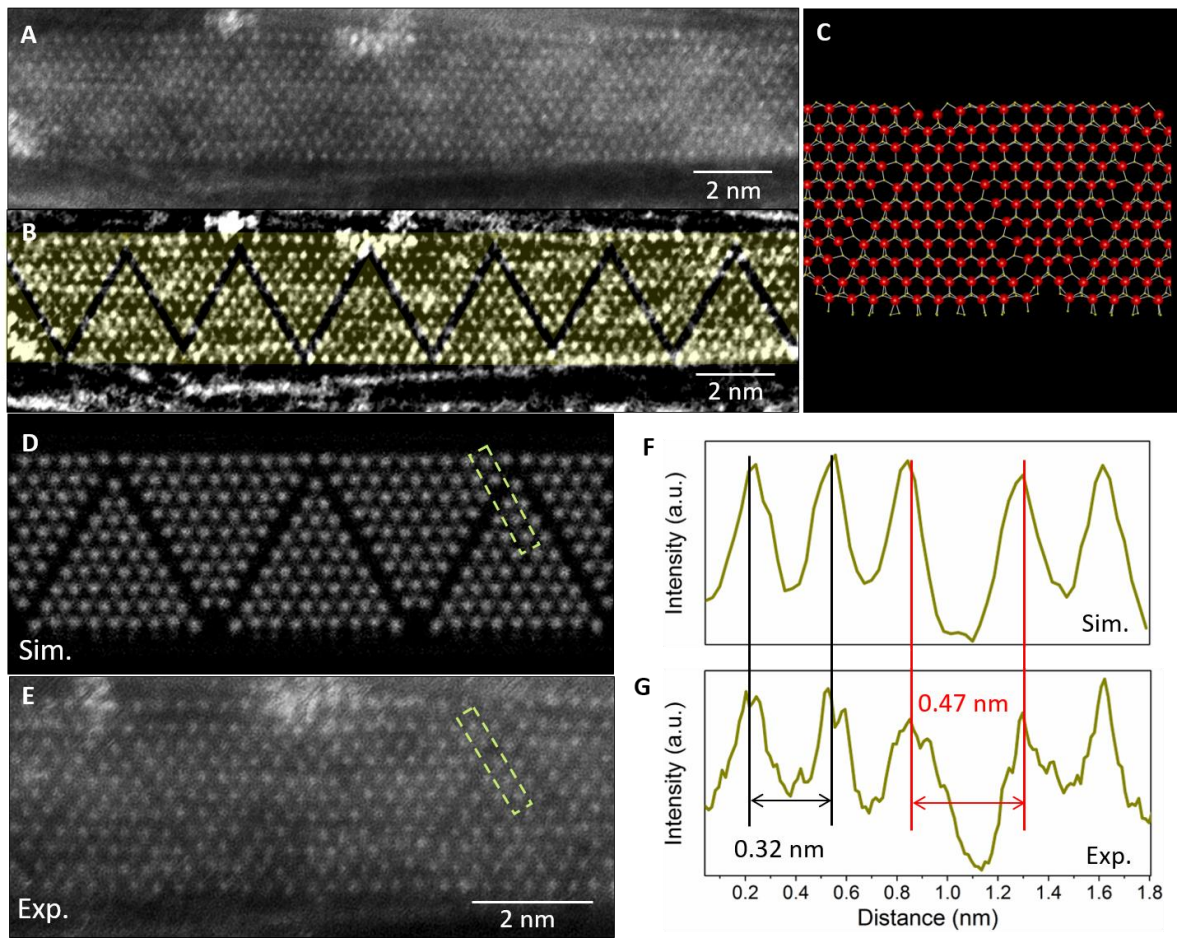

**Fig. 3. Periodic superstructure in TaS₂ nanoribbons.** (**A**) ADF-STEM image and (**B**) its filtered version showing presence of zig-zag like atomic superstructure. The superstructure is identified as defect line arrays, with lines of missing S atoms. (**C**) Calculated atomic structure of defect line arrays in monolayer TaS₂ nanoribbons. (**D**) Simulated an (**E**) experimental STEM images of defect line arrays. Line intensity scans of (**F**) the simulated and (**G**) experimental images across the defect boundary along the dashed green boxes.

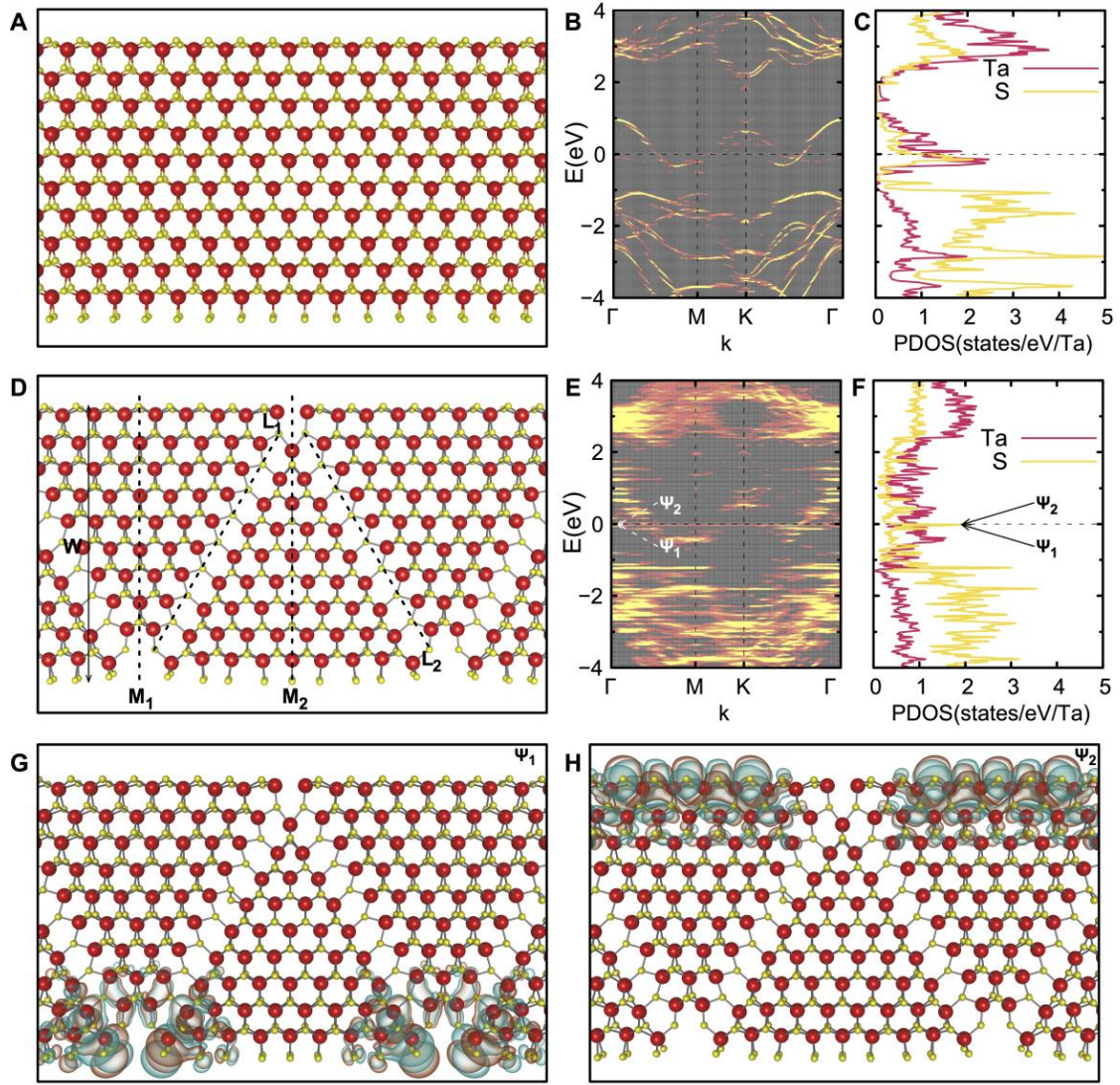

**Fig. 4. Atomic and electronic structure of TaS₂ NRs with and without defect line arrays. (A-C)** The atomic and electronic structures of the 2H-NR of $W$=2.99 nm without the zigzag defect are shown. The structure does have CDW distortions, but the distortion amplitude is too small to be seen by eye in (A). **(A)** shows the atomic structure in planar view, **(B)** the electronic band structure, and **(C)** the PDOS of the NR. **(D-H)** The atomic and electronic structures of 2H-NR of $W$=3.08 nm with zigzag defect array are shown. **(D)** The atomic structure of the NR is shown in the planar view, where the zigzag boundaries of S vacancies are represented by black dashed lines denoted as $L_1$ and $L_2$, and the mirror planes are represented by vertical black dashed lines denoted as $M_1$ and $M_2$. **(E)** The electronic band structure and **(F)** the PDOS of the NR. The localized edge states are marked by arrows and denoted as $\psi_1$ and $\psi_2$. **(G-H)** The real-space

wavefunctions of the localized edges states $\psi_1$ and $\psi_2$. The isosurfaces for the positive and negative values of the real-space wavefunctions are shaded in cyan and orange, respectively. In the atomic structures, Ta and S atoms are represented by red and yellow spheres, respectively. The bands in (**B**) and (**E**) are unfolded with respect to the unit-cell of the primitive 2H-ML. The density of states projected to Ta and S atoms are represented by red and yellow lines, respectively. The Fermi energy is set to zero.

**Table I. Structural and energetic properties of TaS$_2$ NRs.** The structure phase, boundary shape, type of defects, symmetry, and the width, $W$, of two TaS$_2$ NRs obtained with DFT calculations. The number of Ta (S) atoms in the unit cell, $N_{Ta}$ ($N_S$), and the ratio of $N_{Ta}$ to $N_S$ atoms, $R$. The Gibbs free energy of formation, $\delta G$, is defined as $E_{NR} + n_{Ta}\mu_{Ta} + n_S\mu_S$, where $E_{NR}$ is the total energy per atom of the NR, $n_{Ta}$ and $n_S$ are the mole fractions of Ta and S atoms, respectively, and $\mu_{Ta}$ and $\mu_S$ are the binding energies per atom of the α-Ta bulk and crown-shaped S$_8$ molecule, respectively.

|  | NR-1 | NR-2 |
| --- | --- | --- |
| Phase | 2H | 2H |
| Boundary shape | Zigzag | Zigzag |
| Defects | S vacancies | S vacancies |
| Symmetry | Mirror | Mirror |
| $W$ (nm) | 3.08 | 3.34 |
| $N_{Ta}$ | 99 | 120 |
| $N_S$ | 198 | 240 |
| $R$ | 2.00 | 2.00 |
| $\delta G$ (eV) | -0.573 | -0.633 |

# Supplementary Information for

# **Ultra-Narrow TaS₂ Nanoribbons**

Jeffrey D. Cain[1,2,3†], Sehoon Oh[1,2†], Amin Azizi[1,3†], Scott Stonemeyer[1,2,3,4], Mehmet Dogan[1,2], Markus Thiel[1], Peter Ercius[5], Marvin L. Cohen[1,2], and Alex Zettl[1,2,3,*]

*[1]Department of Physics, University of California at Berkeley, Berkeley, CA 94720, USA*

*[2]Materials Sciences Division, Lawrence Berkeley National Laboratory, Berkeley, CA 94720, USA*

*[3]Kavli Energy NanoSciences Institute at the University of California at Berkeley and the Lawrence Berkeley National Laboratory, Berkeley, CA 94720, USA*

*[4]Department of Chemistry, University of California at Berkeley, Berkeley, CA 94720, USA*

*[5]The Molecular Foundry, Lawrence Berkeley National Laboratory, Berkeley, CA 94720 USA*

†These authors contributed equally

*e-mail: azettl@berkeley.edu

**This PDF file includes:**

Materials and Methods

Supplementary Text

Figs. S1 to S4

**Materials and Methods**

**Materials Synthesis***:* Multi-walled carbon nanotubes (CNTs) (Cheap tubes) were opened *via* oxidation at high temperature. They were heated to 515 °C for 1 hour in air, before the filling step. For filling with $TaS_2$ NRs, the opened CNTs, along with elemental sulfur and tantalum powders were sealed under high vacuum ($10^{-6}$ Torr) in 2 mm diameter quartz ampules and held at 760 °C for 1 week, then to cooled room temperature naturally.

**Materials Characterization:** The filled CNTs are dispersed in isopropyl alcohol by bath sonication for 1 hour and drop-cast onto lacey carbon transmission electron microscope grids for imaging. Transmission electron microscope imaging was carried out at 80 keV on a JEOL 2010 microscope. Atomic-resolution STEM imaging is completed at the National Center for Electron Microscopy on TEAM 0.5 which is a Titan 80-300 with an ultra-twin pole piece gap, DCOR probe aberration corrector and was operated at 80 kV and semi-convergence angle of 30 mrad. Images were acquired using the ADF-STEM detector with an inner angle of 60 mrad and a beam current of approximately 70 pA. An aberration-corrected FEI Titan3 (60−300) equipped with a SuperX EDS system at 80 kV was also used for the imaging and spectroscopy. Elemental mapping was performed in the STEM mode at 80 kV with a 7 min acquisition time.

STEM simulations were done using Prismatic's prism algorithm with parameters that matched the experiments. (*1*) In detail for each simulation, we used a 52.92 Å square simulation box, 80 kV accelerating voltage, 30 mrad convergence semi-angle, 0.04 Å/pixel sampling, 1 Å slice size, interpolation factor 4 and 50 frozen phonon calculations. After completion, the simulation data was convolved with a 1.3 Å source size to match the contrast seen in the experiment images. Noise was added following Poisson counting statistics to match the 70 pA experimental beam current. This allowed us to interpret the positions of atoms based on the approximate Z-contrast in the images and to compare the projection images to the DFT simulated structures.

**Computational Methods:** We use the generalized gradient approximation (*2*), norm-conserving pseudopotentials (*3*), and localized pseudo-atomic orbitals for the wavefunction expansion as implemented in the SIESTA code (*4*). The spin–orbit interaction is considered using fully relativistic j-dependent pseudopotentials (*5*) in the l-dependent fully-separable nonlocal form using additional Kleinman–Bylander-type projectors (*6*). We use a $64\times64\times18$ Monkhorst–Pack $k$-point mesh for 2H-bulk, a $64\times64\times36$ mesh for 1T-bulk, a $64\times64\times1$ mesh for 1H- and 1T-MLs, a $1\times64\times1$ mesh for 1H- and 1T-NRs without boundaries, and a $1\times9\times1$ mesh for 1H- and 1T-NRs with boundaries. Real-space mesh cut-off of 1000 Ry is used for all of our calculations. The van der Waals interaction is evaluated using the DFT-D2 correction (*7*). Dipole corrections are included to reduce the fictitious interactions between layers generated by the periodic boundary condition in our supercell approach (*8*).

**Encapsulation in CNT:** To investigate the effects of encapsulation of the $TaS_2$ nanoribbons (NRs) inside the carbon nanotubes (CNTs), separately relaxed atomic positions of $TaS_2$-NRs isolated in vacuum and empty CNTs are used. Further relaxations are not performed. We then calculate the charge transfers between NRs and CNTs. The calculated charge transfer ($q$) from CNTs to $1H$-$TaS_2$ NRs is 0.0192 $e$/Ta atom, where $e$ is the electron charge. Note that $q$ is much smaller than those for the transition metal tricalcogenides such as $NbSe_3$ (*9*), $HfTe_3$ (*10*), and $Hf_2Te_9$ molecular chain (*11*).

**Magnetic Moments:** We perform fully magnetic calculations with spin–orbit interaction. We start with various magnetic moments in the self-consistent calculation, but find no significant magnetic moment in any of the converged calculation results. After carefully checking the magnetic moments, we perform self-consistent calculations again assuming the time-reversal symmetry.

**Topological Properties:** Because of the presence of the mirror symmetry in $1H\text{-}TaS_2$ NRs combined with the time-reversal symmetry, the Zak phases of the bands are quantized (*12*). For a $TaS_2$-NR with a finite total gap, which is not the case in this study but might be the case with smaller width, the topological invariance of the NR can be tuned by controlling the charge state using doping or gating as reported in recent studies (*10*, *13*).

**References**

1.  C. Ophus, A fast image simulation algorithm for scanning transmission electron microscopy. *Adv. Struct. Chem. Imaging.* **3**, 13 (2017).

2.  J. P. Perdew, K. Burke, M. Ernzerhof, Generalized gradient approximation made simple. *Phys. Rev. Lett.* **77**, 3865–3868 (1996).

3.  N. Troullier, J. L. Martins, Efficient pseudopotentials for plane-wave calculations. *Phys. Rev. B.* **43**, 1993–2006 (1991).

4.  J. M. Soler, E. Artacho, J. D. Gale, A. García, J. Junquera, P. Ordejón, D. Sánchez-Portal, "The SIESTA method for ab initio order-N materials simulation" (2002).

5.  G. B. Bachelet, M. Schlüter, Relativistic norm-conserving pseudopotentials. *Phys. Rev. B.* **25**, 2103–2108 (1982).

6.  G. Theurich, N. A. Hill, Self-consistent treatment of spin-orbit coupling in solids using relativistic fully separable ab initio pseudopotentials. *Phys. Rev. B - Condens. Matter Mater. Phys.* **64**, 073106 (2001).

7.  S. Grimme, Semiempirical GGA-type density functional constructed with a long-range dispersion correction. *J. Comput. Chem.* **27**, 1787–1799 (2006).

8.  M. L. Cohen, M. Schlüter, J. R. Chelikowsky, S. G. Louie, Self-consistent pseudopotential method for localized configurations: Molecules. *Phys. Rev. B.* **12**, 5575–5579 (1975).

9.  T. Pham, S. Oh, P. Stetz, S. Onishi, C. Kisielowski, M. L. Cohen, A. Zettl, Torsional instability in the single-chain limit of a transition metal trichalcogenide. *Science.* **361**, 263–266 (2018).

10. S. Meyer, T. Pham, S. Oh, P. Ercius, C. Kisielowski, M. L. Cohen, A. Zettl, "Metal-insulator transition in quasi-one-dimensional HfTe3 in the few-chain limit," (available at https://arxiv.org/ftp/arxiv/papers/1903/1903.00464.pdf).

11. T. Pham, S. Oh, S. Stonemeyer, B. Shevitski, J. D. Cain, C. Song, P. Ercius, M. L. Cohen, A. Zettl, Emergence of Topologically Nontrivial Spin-Polarized States in a Segmented Linear Chain. *Phys. Rev. Lett.* **124**, 206403 (2020).

12. J. Zak, Berrys phase for energy bands in solids. *Phys. Rev. Lett.* **62**, 2747–2750 (1989).

13. T. Cao, F. Zhao, S. G. Louie, Topological Phases in Graphene Nanoribbons: Junction States, Spin Centers, and Quantum Spin Chains. *Phys. Rev. Lett.* **119**, 076401 (2017).

**Figures:**

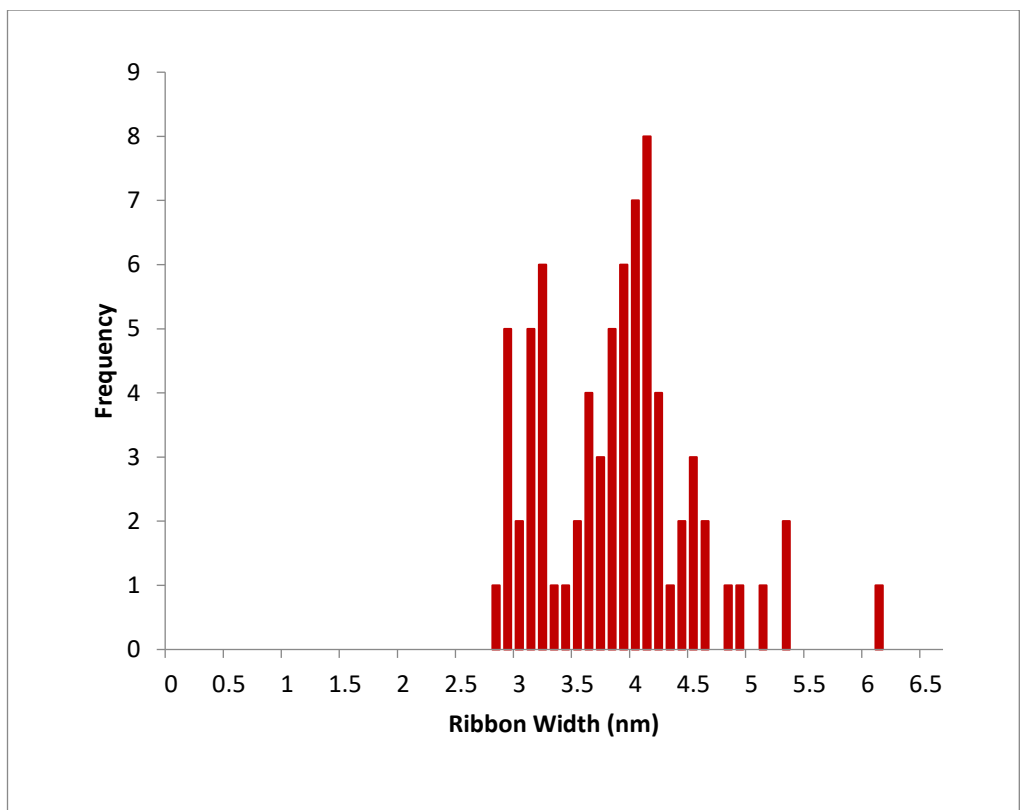

Fig. S1. Histogram of TaS$_2$ nanoribbon widths.

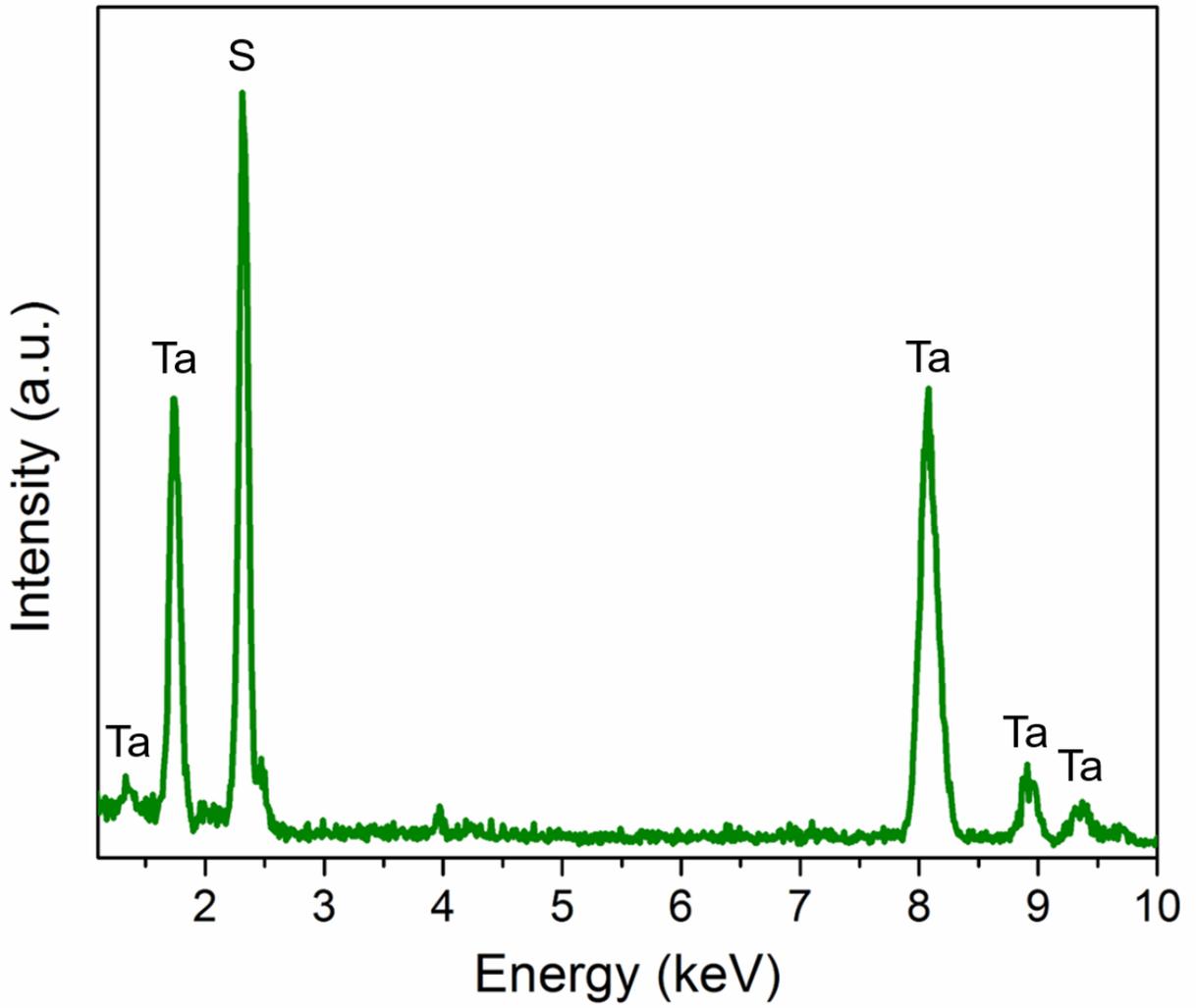

**Fig. S2. Energy dispersive x-ray spectroscopy (EDS).** EDS Spectrum.

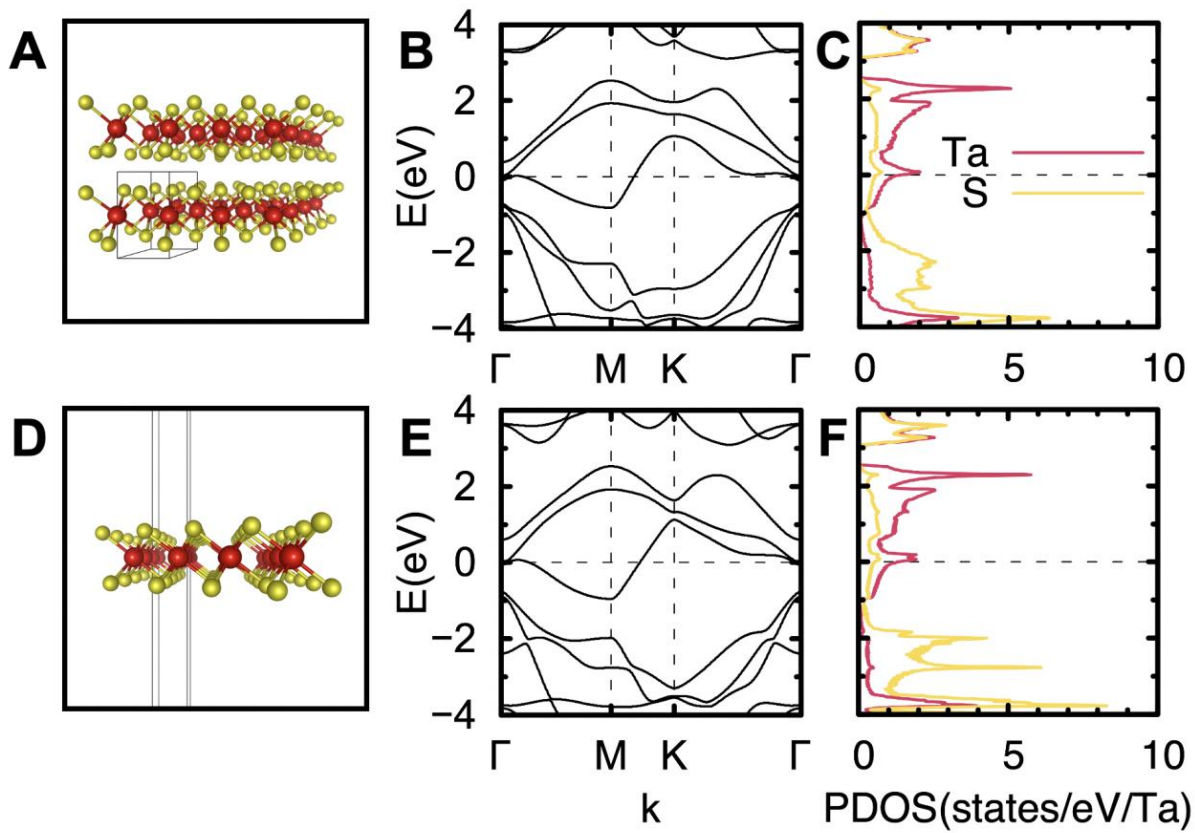

**Fig. S3. Atomic and electronic structures of the TaS₂ bulk and monolayer (ML)**. The atomic and electronic structures of **(A-C)** 2H-bulk, and **(D-F)** 1H-ML are shown. In the 1st column, the atomic structures obtained from density functional theory (DFT) calculations are shown where Ta and S atoms are represented by red and yellow spheres, respectively, and the unit-cell is represented by black frames. In the 2nd column, the electronic band structures are shown where the Fermi energy is set to zero. In the 3rd column, the projected density of states (PDOS) are shown, where the density of states projected to Ta and S atoms are represented by red and yellow lines, respectively.

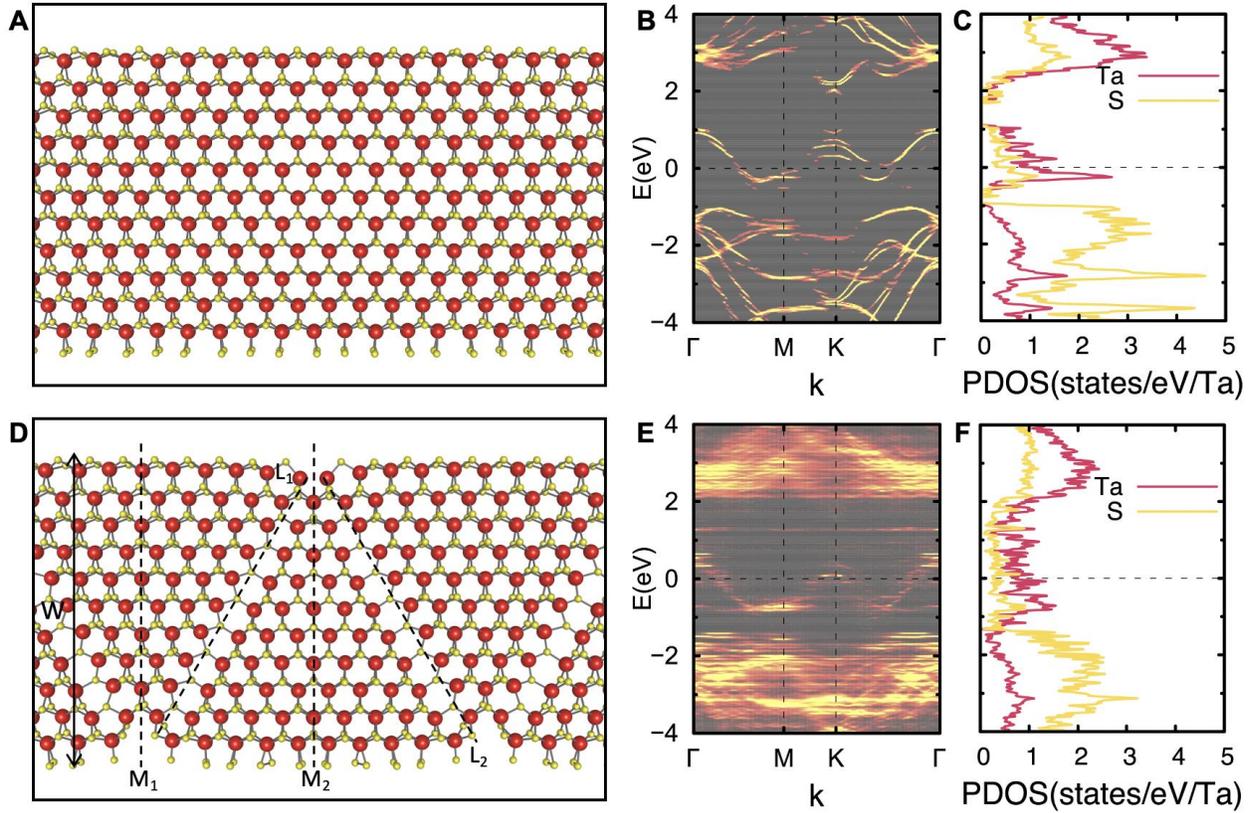

**Fig. S4. Atomic and electronic structure of TaS₂ NRs with and without defect arrays.** (**A**-**C**) The atomic and electronic structures of the 1H-NR of $W$=3.29 nm without the zigzag defect are shown. (**A**) The atomic structure in planar view, (**B**) the electronic band structure and (**C**) the PDOS of the NR are shown. (**D**-**H**) The atomic and electronic structures of 1H-NR of $W$=3.34 nm with zigzag defect array are shown. (**D**) The atomic structure of the NR is shown in the planar view, where the zigzag boundaries of S vacancies are represented by the dashed lines denoted as $L_1$ and $L_2$, and the mirror planes are represented by vertical dashed lines denoted as $M_1$ and $M_2$. (**E**) The electronic band structure and (**F**) the PDOS of the NR are shown. In the atomic structures, Ta and S atoms are represented by red and yellow spheres, respectively. The bands in (**B**) and (**E**) are unfolded with respect to the unit-cell of the primitive 1H-ML. The density of states projected to Ta and S atoms are represented by red and yellow lines, respectively. The Fermi energy is set to zero.